\documentclass[a4paper]{article}

\usepackage{INTERSPEECH2022}
\usepackage{amsmath,graphicx, amsfonts}
\usepackage{multirow}
\usepackage{xcolor}
\usepackage{soul}
\usepackage{enumitem}
\usepackage[flushleft]{threeparttable}
\PassOptionsToPackage{hyphens}{url}\usepackage{hyperref}
\hypersetup{colorlinks=true}
\usepackage{changes}
\usepackage{todonotes}
\usepackage{algpseudocode}
\usepackage{algorithm}

\newcommand{\etal}{\textit{et al.}}
\newcommand{\eg}{\textit{e.g.}}
\newcommand{\ie}{\textit{i.e.}}

\title{Personal VAD 2.0: Optimizing Personal Voice Activity Detection \\ for On-Device Speech Recognition}
\name{Shaojin Ding, Rajeev Rikhye, Qiao Liang, Yanzhang He, \\ Quan Wang, Arun Narayanan, Tom O'Malley, Ian McGraw}
\address{Google LLC, USA}
\email{\{shaojinding, rvrikhye, wildstone, yanzhanghe, quanw\}@google.com}

\begin{document}

\ninept

\maketitle
\begin{abstract}

\noindent
Personalization of on-device speech recognition (ASR) has seen explosive growth in recent years, largely due to the increasing popularity of personal assistant features on mobile devices and smart home speakers. In this work, we present Personal VAD 2.0, a personalized voice activity detector that detects the voice activity of a target speaker, as part of a streaming on-device ASR system. Although previous proof-of-concept studies have validated the effectiveness of Personal VAD, there are still several critical challenges to address before this model can be used in production: first, the quality must be satisfactory in both enrollment and enrollment-less scenarios; second, it should operate in a streaming fashion; and finally, the model size should be small enough to fit a limited latency and CPU/Memory budget. To meet the multi-faceted requirements, we propose a series of novel designs: 1) advanced speaker embedding modulation methods; 2) a new training paradigm to generalize to enrollment-less conditions; 3) architecture and runtime optimizations for latency and resource restrictions. Extensive experiments on a realistic speech recognition system demonstrated the state-of-the-art performance of our proposed method.

\end{abstract}
\section{Introduction}

\noindent
Personal Voice Activity Detection (VAD)~\cite{ding2020personal} aims to detect the voice activity of a target speaker (or multiple target speakers) at the frame level. By rejecting frames which do not contain target user speech, Personal VAD is able to significantly reduce computational resources and also improves speech recognition. As a result, Personal VAD is an  indispensable module for speech processing systems within personalized devices and services, such as smartphones, smart home speakers, and tablets.

Several studies~\cite{ding2020personal, medennikov2020target, he2021target, makishima2021enrollment, jayasimha2021personalizing} have demonstrated the viability of Personal VAD. Ding \etal \cite{ding2020personal} first proposed the idea of a Personal VAD through a proof-of-concept study on the frame-wise VAD accuracy. Following this, Medennikov \etal~\cite{medennikov2020target} and He \etal~\cite{he2021target} investigated the use of Personal VAD as a speaker diarization technique in the “cocktail-party problem” setting~\cite{cherry1953some} with one or more target speakers. Makishima \etal~\cite{makishima2021enrollment} proposed an enrollment-less training scheme for Personal VAD to avoid the need for an enrollment utterance during training. In addition, the authors of ~\cite{jayasimha2021personalizing} extended Personal VAD for the purposes of start/end of speech detection, being more flexible in an ASR system.

However, there are still several critical challenges blocking the use of Personal VAD in production environments, especially on-device speech recognition systems~\cite{sainath2020streaming, he2019streaming, macoskey2021amortized, joshi2020transfer, wang2020low}. Once deployed, Personal VAD will be an ``always running" system that replaces standard VAD. Therefore, it should achieve satisfactory performance in both enrollment and enrollment-less scenarios. With enrollment, it should minimize the insertion error caused by non-target speech while avoiding deleting target speech. Without enrollment, it should perform at least as well as a standard VAD. From the perspective of deployment, Personal VAD should operate in a streaming fashion and have a small model size to have minimal latency and CPU/memory impacts.

The conventional Personal VAD~\cite{ding2020personal, medennikov2020target, he2021target, makishima2021enrollment, jayasimha2021personalizing} currently does not meet these requirements. First, the speaker embedding modulation approaches are usually based on concatenating the acoustic features with the speaker embedding, which we have found to result in sub-optimal VAD and ASR performance (Section~\ref{sec:ablation}). Second, conventional Personal VAD models assume an application scenario with at least one enrolled target speaker, which does not work when no enrolled speaker is present. Lastly, previous works are mostly proof-of-concept studies that do not consider any runtime optimizations, which is unfeasible for production environments.

To address the problems, we propose Personal VAD 2.0, an optimized Personal VAD model for on-device speech recognition. Our main contributions are outlined below:

\begin{itemize}[leftmargin=*]
    \item We propose two advanced methods for speaker embedding modulation, through a feature-wise transform (FiLM) layer~\cite{dumoulin2018feature} or through a speaker pre-net, instead of naively concatenating them to the input acoustic features. 
    \item We propose a novel model training paradigm to extend personal VAD to behave as a standard VAD under enrollment-less conditions. During training, we randomly replace the target speaker embeddings with a zero vector and modify the labels of non-target speaker speech to target speaker speech. 
    \item We also investigate model architecture and runtime optimizations, including using a streaming-friendly and more powerful Conformer backbone~\cite{gulati2020conformer}, and quantizing the model to 8-bit integer format to reduce model size to meet the strict on-device production requirements.
    \item We for the first time conduct experiments on an on-device ASR system with realistic speech traffic to evaluate the effectiveness of personal VAD models. Results show that our proposed model can significantly reduce ASR insertion errors in the enrollment scenario, while retaining the same performance as standard VAD in the enrollment-less scenario.
\end{itemize}

\section{Methods}

In this section, we start with a quick recap of the conventional Personal VAD model. Following this, we introduce our proposed methods of improving speaker embedding modulation, generalizing to both enrollment and enrollment-less conditions, and architecture as well as runtime optimizations.

\vspace{-2pt}
\subsection{Recap: Personal VAD model}
\label{sec:recap}
\vspace{-3pt}

In conventional Personal VAD systems, users need to go through an enrollment process, where a pre-trained text-independent speaker recognition model~\cite{wan2018generalized} computes the target user's d-vector speaker embeddings from the target user’s recordings, to encode their voice characteristics.

\begin{figure}
	\centering
	\includegraphics[width=0.9\columnwidth]{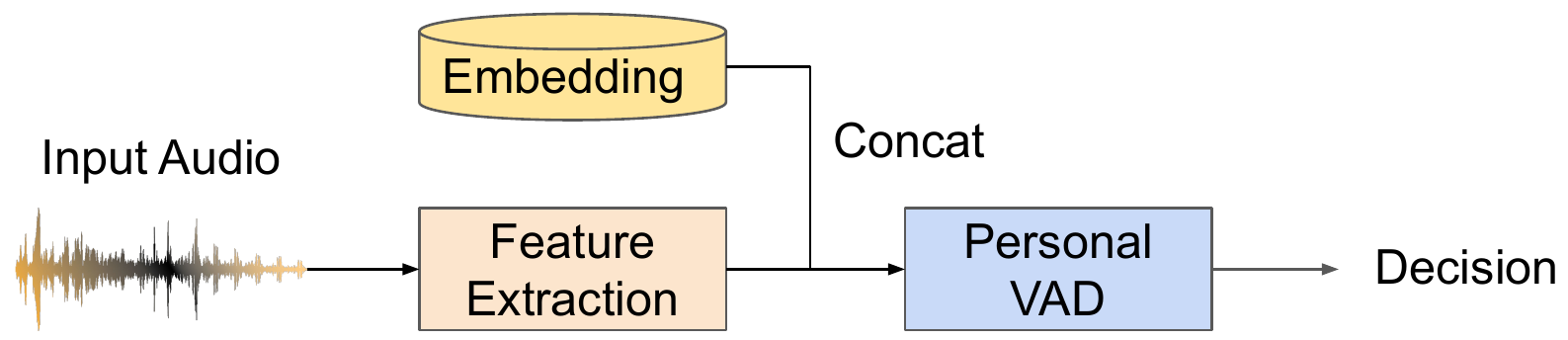}
	\caption{Conventional Personal VAD model. The model takes acoustic features and speaker embedding as the inputs and produces a frame-wise prediction.}
	\label{fig:conv_pvad}
	\vspace{-10pt}
\end{figure}

The diagram of a conventional Personal VAD is shown in Figure~\ref{fig:conv_pvad}. The model first extracts acoustic features $\mathbf{x}\in\mathbb{R}^{T\times D}$ from input audio, where $T$ and $D$ denote the sequence length and feature dimension, respectively, and $\mathbf{x}_t$ represents the $t$-th frame of acoustic features. Each $\mathbf{x}_t$ is then concatenated with the speaker embedding from the target speaker $\mathbf{e^\mathrm{target}}$:

\vspace{-5pt}
\begin{equation}
\label{eq:et}
  \hat{\mathbf{x}}_t = [\mathbf{x}_t, \mathbf{e^\mathrm{target}}] .
\end{equation}

\noindent
The model consumes the concatenated features as inputs and produces frame-wise decision among target speaker's speech (\texttt{tss}), non-target speaker's speech (\texttt{ntss}), and non-speech (\texttt{ns}):

\vspace{-12pt}
\begin{equation}
\label{eq:pvad}
  \mathbf{p}_t = \mathrm{PVAD}(\hat{\mathbf{x}_t}) ,
\end{equation}

\noindent
where $\mathbf{p}_t=[p_t^{\tt tss}, p_t^{\tt ntss}, p_t^{\tt ns}]$ correspond to the posteriors of the three classes. Typically, previous studies implement the model $\mathrm{PVAD}$ with a couple of Bidirectional LSTM (BLSTM) layers (\eg, three-layer BLSTM in~\cite{ding2020personal, medennikov2020target, makishima2021enrollment}).

\subsection{Improving speaker embedding modulation}

As described in Section~\ref{sec:recap}, vanilla Personal VAD models concatenate target speaker embedding with the acoustic feature to incorporate speaker specific information. However, acoustic features and speaker embeddings represent very different information, and they are extracted through entirely separated processes, leading to different distributions and magnitudes. Therefore, simply concatenating them and feeding them to the same layers may significantly limit the model capacity. To address this issue, we propose two novel speaker embedding modulation approaches: 1) through a FiLM layer, and 2) through a speaker pre-net.

\vspace{-2pt}
\subsubsection{Speaker embedding modulation through FiLM}
\vspace{-3pt}

A FiLM layer~\cite{perez2018film} applies a feature-wise affine transformation to its inputs, including scaling and shifting operations. This affine transformation generalizes concatenation-, biasing-, and scaling-based conditioning operators, which has been shown to be more expressive in learning conditional representations than using either of them individually~\cite{perez2017learning, johnson2017clevr}.

\begin{figure}
	\centering
	\includegraphics[width=\columnwidth]{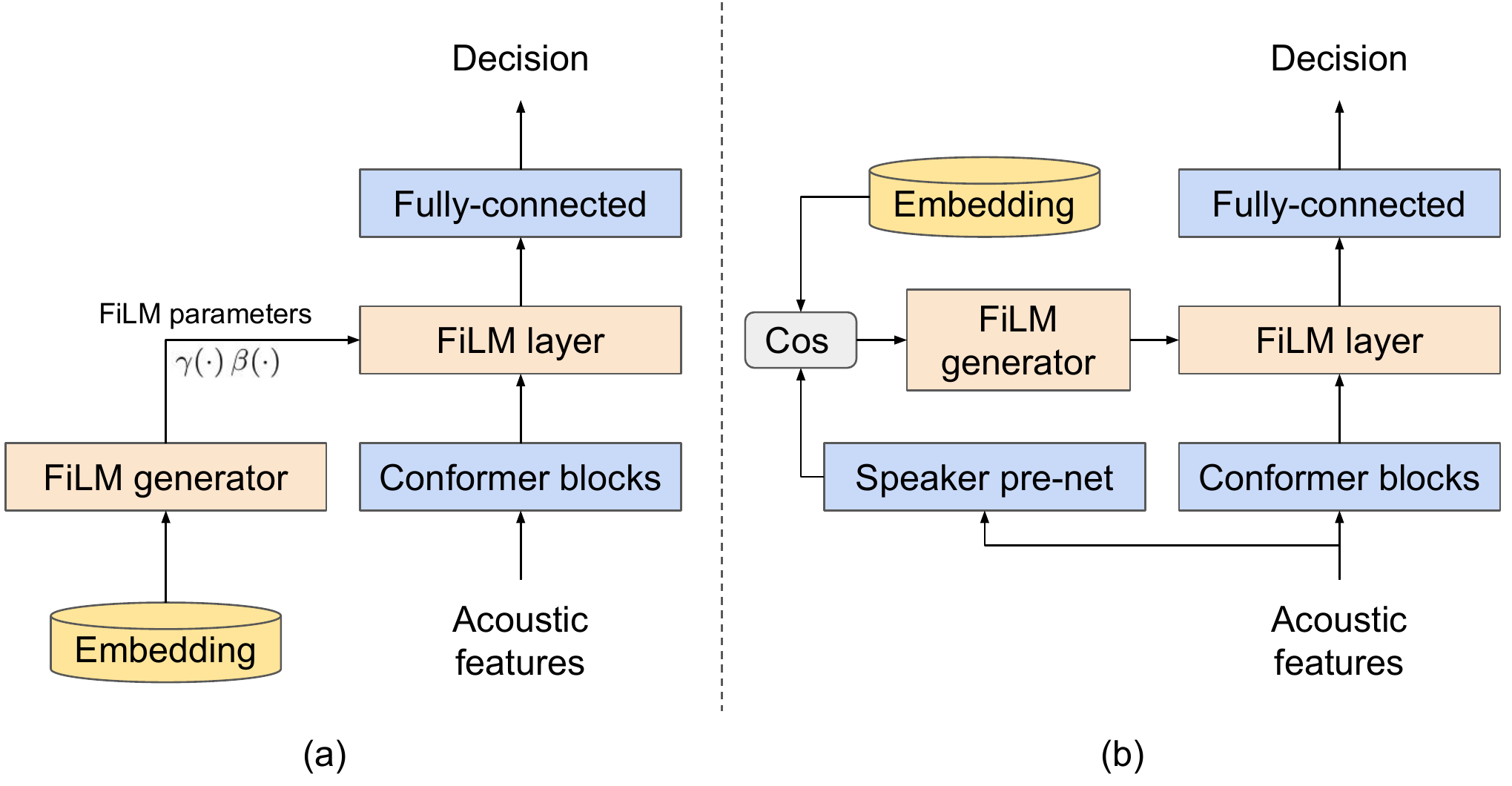}
	\caption{(a) Personal VAD model with FiLM layer. $\gamma(\cdot)$ and $\beta(\cdot)$ are the scaling and shifting vectors of FiLM respectively. (b) Personal VAD model with speaker pre-net.}
	\label{fig:film}
	\vspace{-20pt}
\end{figure}

Figure~\ref{fig:film}(a) shows the diagram of Personal VAD model with FiLM layer. Suppose $\mathbf{h} = \mathrm{Conformer}(\mathbf{x})$ is the input to FiLM layer. The FiLM generator first takes the target speaker embedding $\mathbf{e^\mathrm{target}}$ as inputs, and produces a scaling vector $\gamma(\mathbf{e^\mathrm{target}})$ and a shifting vector $\beta(\mathbf{e^\mathrm{target}})$ (\ie, FiLM parameters), with the same dimensions as $\mathbf{h}$. Following this, the input $\mathbf{h}$ is scaled and shifted by the corresponding vectors:

\vspace{-5pt}
\begin{equation}
\label{eq:film}
  \mathrm{FiLM}(\mathbf{h}) = \gamma(\mathbf{e^\mathrm{target}})\cdot \mathbf{h} + \beta(\mathbf{e^\mathrm{target}})
\end{equation}

\noindent
The output of the FiLM layer is passed to the final fully-connected classifier.

\vspace{-3pt}
\subsubsection{Speaker embedding modulation through speaker pre-net}
\vspace{-3pt}

Alternatively, instead of directly conditioning speaker embeddings to the network, we propose adding a speaker pre-net to extract the speaker information from the acoustic features and produce a learned embedding that has the same dimensions as the speaker embedding. Following this, we compute the cosine similarity between the learned embedding and the target speaker embedding, which is then used for conditioning. Compared to direct speaker embeddings conditioning, this will provide more discriminative information to the classifier, which assists the model to better distinguish different speaker and provide more accurate decisions.

Figure~\ref{fig:film}(b) shows the diagram of Personal VAD with speaker pre-net. Formally, using our prior notations, the speaker pre-net consumes acoustic features $\mathbf{x}$ and produces a fixed-length embedding for each frame $\mathbf{e}^{\mathrm{prenet}} \in \mathbb{R}^{T\times D_e}$:

\vspace{-5pt}
\begin{equation}
\label{eq:prenet}
  \mathbf{e}^{\mathrm{prenet}} = \mathrm{PreNet}(\mathbf{x})
\end{equation}

\noindent
where $D_e$ is the dimension of the pre-net embedding (equal to that of speaker embedding). Then, we compute the cosine similarity score $\mathbf{s}\in \mathbb{R}^{T}$ between the two embeddings:

\vspace{-5pt}
\begin{equation}
\label{eq:cos}
  \mathbf{s} = \cos(\mathbf{e}^{\mathrm{prenet}}, \mathbf{e}^{\mathrm{target}})
\end{equation}

\noindent
Lastly, we modulate the output of the conformer blocks with the cosine similarity score via FiLM:

\vspace{-5pt}
\begin{equation}
\label{eq:film_cos}
  \mathrm{FiLM}(\mathbf{h}) = \gamma(\mathbf{s})\cdot \mathbf{x} + \beta(\mathbf{s})
\end{equation}
As a straightforward extension, we also investigated modulating both speaker embedding and cosine similarity with FiLM (concatenating the two and then feed to FiLM layer) to consider both generative and discrimination speaker-related clues, which we will examine in the experiments.

\vspace{-2pt}
\subsection{Training the model for both enrollment and enrollment-less conditions}
\vspace{-3pt}

In practical on-device ASR use cases, the user may choose to either proceed with or skip the enrollment process. As an "always-running system", the Personal VAD model should be able to achieve satisfactory performance in both conditions.
To achieve this, we propose a simple but effective training paradigm to guarantee satisfactory performance for both enrollment and enrollment-less conditions, as shown in Alg.~\ref{alg:training}. This guarantees the model learning the standard VAD behaviors while there is no speaker embedding. In practice, we set $p_0=0.2$, which we found to provide reasonable performance for both conditions. Additionally, we add datasets that have no enrollment utterance to the model training, paired with a zero speaker embedding for each utterance. During inference, we also feed an zero speaker embedding to the model under the enrollment-less condition.

\vspace{-10pt}
{\begin{algorithm}[H]
    \small
    \caption{Training the model for both enrollment and enrollment-less conditions.}
    \begin{algorithmic}[1]
    \State During each training epoch, sample a subset of training utterances with probability $p_0$
    \For{each sampled training utterance}
        \State Set the speaker embedding to zero vector: $\mathbf{e}^{\mathrm{target}}=\overrightarrow{\mathbf{0}}$
        \State Replace ground-truth labels of \texttt{ntss} to \texttt{tss}
    \EndFor
    \end{algorithmic}
    \label{alg:training}
\end{algorithm}}
\vspace{-13pt}

\vspace{-3pt}
\subsection{Conformer backbone: Streaming support \& accuracy improvement}
\vspace{-4pt}

Transformer models~\cite{vaswani2017attention} have been shown to substantially improve almost all speech modeling tasks over LSTMs in recent years~\cite{dong2018speech, gulati2020conformer, li2019neural}. Consequently, we investigate the use of Conformer backbone, an advanced Transformer architecture, for Personal VAD. The vanilla transformer uses the full sequence as the attention context, which leads to a very high latency in streaming systems that is similar to BLSTMs. To make the model stream-able and maximally reduce the latency, we set the model to have a limited left-context and no right-context, following~\cite{moritz2020streaming, li2021better}.

\vspace{-3pt}
\subsection{Model Quantization}
\vspace{-4pt}

Due to the limited resources on mobile devices, we adopt a runtime optimization for the on-device inference of Personal VAD. Instead of directly running TensorFlow graph with 32-bit float weights, we quantize the model weights to have 8-bit integers using dynamic range quantization~\cite{alvarez2016efficient, shangguan2019optimizing} and serialize the
models to the TensorFlow Lite format. The quantized model has only 1/4 model size, and more importantly, achieves significant speedup with optimized hardware instructions of integer arithmetic, improving both memory usage and latency for Personal VAD.

\vspace{-2pt}
\section{Experimental setup}

\vspace{-2pt}
\subsection{Datasets}
\vspace{-3pt}

We conduct our experiments using a vendor-collected dataset of realistic speech queries. The training set contains 2.6 million utterances ($\sim$1,600 hours) from 6,923 speakers. As suggested by~\cite{ding2020personal}, there is no existing datasets that is ideal to Personal VAD evaluations, where each utterance contains natural speaker turns and it contains enrollment utterances for each individual speaker. To simulate the conversational speech, we create a set of synthetic data based on the vendor-collected dataset, following~\cite{ding2020personal}. We first concatenate utterances from multiple speakers, and then we randomly select one of the speakers as the target speaker in the concatenated utterance. Accordingly, we extract the target speaker embedding from the utterance in an enrollment subset. In addition, we also apply two data augmentation techniques, MTR~\cite{Kim2017} and SpecAug~\cite{park2019specaugment}, on our datasets to avoid domain overfitting and mitigate concatenation artifacts. Most of the models in the experiments are trained with concatenated dataset alone. Only when training the model for both enrollment and enrollment-less conditions, we used an additional internal training set~\cite{sainath2020streaming} consisting of 35 million English \textit{audio-text pairs} ($\sim$27,500 hours) from multiple domains including YouTube and anonymized voice search traffic. The internal training set is only used for enrollment-less condition (where we do not perform any speaker-related operations), and our data handling abides by \textit{Google AI Principles}~\cite{googleaiprinciples}. We use the same datasets in our baseline standard VAD training for a fair comparison. 

The test set of the vendor-collected dataset of realistic speech queries comprises 194,890 utterances from 1,241 speakers. These utterances are very similar to actual speech query traffic and can contain various ambient noise. As a result, we do not add extra reverberation or noise to it during evaluations.  The non-concatenated test set shows the performance when only the target speaker is talking. Similarly, we create a concatenated test set to evaluate the performance of Personal VAD in conversational speech. To reduce the evaluation cost, we randomly samples a subset of 4,000  utterances from the original test set and the concatenated test set, respectively. In the evaluations under enrollment-less condition, we also use 14,000 Voice-search utterances (VS). 

\vspace{-4pt}
\subsection{Implementation details}
\vspace{-4pt}

We use the same frontend feature as ~\cite{sainath2020streaming}, a 128-dimensions log-Mel filterbank energies extracted with 32ms window and 10ms shift. We stack the features from 4 contiguous frames (512-dimensional) and subsample the sequence by a factor of 3. The LSTM-based models have 3 uni-directional LSTM layers with 256 units and a final fully-connected layer as the classifier. Our Conformer-based models have 4 Conformer layers in the block, each having a dimension of 64, attention head of 8, causal $7\times 7$ convolution kernel, and 31 left-context. The speaker pre-net has two Conformer layers with the same hyper-parameters.

During inference, we compare the posterior of target speaker speech $p_t^{\tt tss}$ to a pre-defined threshold ($0.1$ in all our experiments) to decide if a speech frame belongs to the target speaker. We discard non-target speech and non-speech frames and only pass target speech frames to downstream ASR model. We use a medium size ($\sim$ 40M parameters) LSTM-based ASR model~\cite{he2019streaming} among all our experiments, which is trained on the realistic voice search traffic. We use word error rate (WER) to measure the ASR performance with Personal VAD models.

\vspace{-4pt}
\section{Results}
\vspace{-2pt}

\begin{table}[t]
\begin{center}
\caption{WERs (\%) of the ablation study on the proposed components under the enrollment condition. Non-Concat test set shows the performance when only the target speaker is talking, while the concatenated test set showing the performance in a conversational speech. Model size is represented in MB. Runtime complexity is measure by FLOPs.}
\vspace{-9pt}
\label{table:ablation}
\resizebox{\columnwidth}{!}{
\begin{tabular}{c|l|cc|cc}
\hline
\multirow{2}{*}{Exp} & \multicolumn{1}{c|}{\multirow{2}{*}{Model}} & \multicolumn{2}{c|}{Speech Query test set} & Size & FLOPs \\
\cline{3-4} & & Non-Concat & Concat & (MB) & (M)  \\
\hline
B0 & Personal VAD~\cite{ding2020personal} & 17.9 & 41.0 & 5.8 & 3.54 \\
\hline
E0 & + Conformer & 15.3 & 31.5 & 2.8 & 9.51 \\
\hline
E1 & \hspace{2mm}+ FiLM d-vector & 11.3 & 29.5 & 2.8 & 9.58 \\
E2 & \hspace{2mm}+ Speaker PreNet (cos) & 11.7 & 27.5 & 4.0 & 9.51 \\
E3 & \hspace{2mm}+ FiLM d-vector \& cos & 11.5 & 27.6 & 4.0 & 9.58\\
\hline
E4 & \hspace{4mm}+ 8-bit quantization & 12.2 & 27.2 & 1.0 & 9.58 \\

\hline

\end{tabular}}
\end{center}
\vspace{-25pt}
\end{table}

\begin{table*}[ht]
\begin{center}
\caption{Overall comparisons to baseline standard VAD and Personal VAD, and the ablation study results of the proposed training paradigm for both enrollment and enrollment-less conditions. Under enrollment-less condition, the WERs on Concat testset are expected to be over 100\%, so we do not show it here for simplicity.}
\vspace{-9pt}
\label{table:overall}
\resizebox{0.7\textwidth}{!}{
\begin{tabular}{c|c|cc|cc|cc}
\hline
\multirow{2}{*}{Exp} & \multirow{2}{*}{Method} & \multicolumn{2}{c|}{Enrollment} & \multicolumn{2}{c|}{Enrollment-less} & \multirow{2}{*}{Size (MB)} & \multirow{2}{*}{FLOPs (M)} \\
 \cline{3-6} & & Non-Concat & Concat & VS & Non-Concat   \\
\hline
B1 & LSTM Standard VAD & N/A & N/A & 7.2 & 11.5 &  1.5 & 3.41 \\
B2 & Conformer Standard VAD & N/A & N/A & 6.9 & 10.1 & 0.7 & 8.77 \\
\hline
B0 & Personal VAD~\cite{ding2020personal} & 17.9 & 41.0 & $\geq$100.0 & $\geq$100.0 & 5.8 & 3.54 \\
E5 & Personal VAD 2.0 & 12.4 & 32.7 & 7.0 & 10.1  & 1.0 & 9.58 \\
\hline
\hline
E4 & w/o our training paradigm & 12.2 & 27.2 & $\geq$100.0 & $\geq$100.0 & 1.0 & 9.58 \\
E5 & w/ our training paradigm & 12.4 & 32.7 & 7.0 & 10.1 & 1.0 & 9.58 \\
\hline
\end{tabular}}
\vspace{-23pt}
\end{center}
\end{table*}

We conducted a series of ablation studies and comparisons to evaluate the performance of the proposed Personal VAD 2.0. Starting from the conventional Personal VAD model in~\cite{ding2020personal}, we incrementally investigate the performance gain (adding one technique at a time) obtained from Conformer backbone, novel speaker embedding modulation methods, model quantization, and the model training paradigm for enrollment and enrollment-less conditions. Finally, we made an overall comparison of the best performing Personal VAD 2.0 model to the baselines to emphasize the effectiveness of our proposed approach.

\vspace{-4pt}
\subsection{Ablation studies}
\vspace{-4pt}
\label{sec:ablation}

\textbf{Conformer backbone.} We first evaluate the efficacy of the Conformer backbone. As shown in Table~\ref{table:ablation}, using Conformer backbone (\textit{E0}) significantly improves both Non-Concat and Concat test sets by 2.6 and 9.5 absolute WERs, respectively. More importantly, the model size of the Conformer model is less than a half of the LSTM model size (\textit{E0}: 2.8MB vs. \textit{B0}: 5.4MB), suggesting that Conformer-based Personal VAD model can achieve much high parameter efficiency.

\noindent
\textbf{Proposed speaker embedding modulation methods.} Next, we evaluate our proposed speaker embedding modulation methods: (\textit{E1}) FiLM, (\textit{E2}) Speaker pre-net (\ie, FiLM cos similarity score), and (\textit{E3}) the combination of both. Table~\ref{table:ablation} shows the performance of the three approaches. Compared to concatenating at the input (\textit{E0}), all three approaches achieve essential performance gains on both Non-Concat ($\geq$3.6) and Concat ($\geq$2.0) test sets, indicating that the proposed modulation approaches have better capacity for speaker-related information. When comparing between the three approaches, we found that they perform similarly on Non-Concat, but \textit{E2} and \textit{E3} achieve further improvement on Concat, indicating the importance of modulating discriminative information to the models.

\noindent
\textbf{Model quantization.} We also evaluate the influence of model quantization in final ASR WERs. We added 8-bit model quantization on the top of \textit{E3}, as shown in system \textit{E4} in Table~\ref{table:ablation}. From the results, we found that the quantization hurts the WER on Non-Concat test set, but the degradation is marginal (0.7). Interestingly, the Concat WER is improved by 0.2, possibly due to the quantization avoid the model overfitting to some local noises. With the 75\% model size reduce and the marginal performance change, model quantization proves its effectiveness and necessity in on-device deployment.

\noindent
\textbf{Training the model for both enrollment and enrollment-less conditions.} Lastly, we explore the model training paradigm that is proposed to guarantee Personal VAD to have reasonable performance under enrollment-less condition. Results are shown in Table~\ref{table:overall}. Under enrollment-less condition, the model with it (\textit{E5}) can achieve 7.0 and 10.1 on VS and Non-Concat test sets, respectively, which suggests that adopting the paradigm can successfully generalize Personal VAD to enrollment-less condition. Surprisingly, compared to \textit{E4}, \textit{E5} has a 4.5 WER increase under enrollment condition. To analyze this observation, we break the WER down to deletion, insertion, and substitution errors (\textit{E4}: 3.7/14.8/8.9 vs. \textit{E5}: 4.7/18.9/9.1). We noticed that the increased WER is mainly due to an increased insertion error rate. A possible explanation is that, with the joint training paradigm, the model is biased to provide a higher posterior for target speaker speech, as we use this dimension for all the speech frames when dealing with enrollment-less data.

\vspace{-4pt}
\subsection{Overall comparisons to baselines and discussions}
\vspace{-4pt}

After verifying the impact of each aspect of our proposed approach, we deliver the comparison results between  the best performing Personal VAD 2.0 model and standard/Personal VAD baselines in Table~\ref{table:overall}. For a fair model size comparison, we also conduct 8-bit quantization to the standard VAD baseline, as they have been widely used in on-device products. Compared with the two standard VAD baselines, Personal VAD 2.0 can significantly reduce the WER (mostly insertion errors) on conversational speech scenarios (Concat test set), while retaining the performance on regular speech queries under either condition. Personal VAD 2.0 has higher FLOPs, due to the use of Conformer backbone. However, it does not have significant impacts to the overall ASR system latency since VAD has much less computations than ASR. More importantly, conformer backbone makes it possible to process several frames in a the same batch, which further speed up model inference and reduces the latency.

Compared with the conventional Personal VAD, Personal VAD 2.0 not only tremendously improves the performance under the enrollment condition but also generalizes well to enrollment-less conditions. More importantly, with the optimizations based on Conformer backbone and 8-bit weight quantization, we developed a light-weight Personal VAD model that can easily fit into any on-device systems. Meanwhile, we still notice an issue of the current Personal VAD 2.0 model -- it performs slightly better on regular speech queries (Non-Concat) under enrollment-less condition than under enrollment condition. A potential reason is that when the target speaker and non-target speaker have very similar voice, the model is more likely to make erroneous predictions. The gap can be reduced by tuning the VAD decision threshold, at the cost of an increasing insertion error rate on conversational speech, resulting in a trade-off to be made according to the deployment scenarios (\eg, for keyword spotting, we need to minimize deletion error, and it is less sensitive to insertion errors).

\vspace{-3pt}
\section{Conclusions}
\vspace{-3pt}

In this study, we have proposed Personal VAD 2.0, an optimized Personal VAD model for on-device speech recognition. Through a series of ablation studies, we evaluate the impact of our novel design choices and runtime optimizations. Specifically, in terms of speaker embedding modulation, we showed that using FiLM and speaker pre-net can significantly improve the model performance than simply concatenation at the input. We also confirmed that our proposed training paradigm can effectively generalize Personal VAD to enrollment-less inference setting, while retaining the performance under enrollment condition. Additionally, we verified that adopting Conformer backbone and adding 8-bit quantization to Personal VAD model can tremendously increase the parameter efficiency. Compared to baseline standard VADs and conventional Personal VAD, Personal VAD 2.0 achieves the state-of-the-art performance in an on-device speech recognition task. Future work will focus on mitigating the performance difference between enrollment and enrollment-less conditions, and investigating multi-user Personal VAD in on-device ASR scenarios.

\bibliographystyle{IEEEtran}

\bibliography{mybib}

\end{document}